# The Artist is Present: Evidence of Artists Residing and Spawning in Text-to-Audio AI

**Guilherme Coelho**
Audio Communication Department
Technische Universität Berlin
*guilherme.coelho@thenewcentre.org*

## Abstract

Text-to-audio (TTA) systems are rapidly transforming music creation and distribution, with platforms like Udio and Suno generating thousands of tracks daily and integrating into mainstream music platforms and ecosystems. These systems, trained on vast and largely undisclosed datasets, are fundamentally reshaping how music is produced, reproduced and consumed. This paper presents empirical evidence that artist-conditioned regions can be systematically *microlocated* through metatag-based prompt design, effectively enabling the spawning of artist-like content through strategic prompt engineering. Through systematic exploration of metatag-based prompt engineering techniques, this research reveals how users can access the distinctive sonic signatures of specific artists, evidencing their inclusion in training datasets. Using descriptor constellations drawn from public music taxonomies, I demonstrate reproducible proximity to artists such as Bon Iver, Philip Glass, Panda Bear and William Basinski. The results indicate stable text–audio correspondences consistent with artist-specific training signals, enabling precise traversal of stylistic microlocations without explicitly naming artists. This capacity to summon artist-specific outputs shows that artists' creative works function as foundational material from which these systems generate new content, often without explicit consent or attribution. In this sense, the artist is indeed present within these generative systems, operating as a rhizomatic foundation from which seemingly novel works emerge**.** Conceptually, the work clarifies how textual descriptors act as navigational cues in high-dimensional representation spaces; methodologically, it provides a replicable protocol for auditing stylistic inducibility. The findings raise immediate questions for governance—attribution, consent, and disclosure standards—and for creative practice, where induced stylistic proximity complicates boundaries between ownership, reproduction, imitation, creative agency and the ethics of algorithmic creation.

## 1    Introduction

Text-to-audio (TTA) AI systems represent a significant paradigm shift in music creation, offering unprecedented capabilities to generate compositions through natural language prompts. Contemporary platforms such as Udio and Suno exemplify this capability, delivering on-demand outputs that often exhibit marked proximity to established genres, compositional idioms, production aesthetics, and, critically, particular artists. Unlike models that disclose salient characteristics of their training data (e.g., Stable Audio), these platforms provide little transparency regarding sources and curation practices, foregrounding questions of intellectual-property governance, data provenance, and creative agency (Stability AI, 2023; Betker et al., 2023). These systems are transforming the music landscape at an unprecedented scale. Industry reporting suggests that Udio alone generates approximately 10 songs per second—over 800,000 daily (Metz, 2024). Recent estimates indicate that around 18% of new tracks uploaded to Deezer are AI-generated (Deezer, 2025), while AI-generated tracks have achieved viral status, exemplified by the Metro Boomin remixed "BBL Drizzy" with 3.3 million streams (Willonius, 2024)[1] and Velvet Sundown's fully AI-generated *Floating on Echoes* album,[2] which has accrued millions of plays on Spotify (Pareño, 2025).[3] Underpinning these phenomena are vast models trained on immense corpora of undisclosed pre-existing artistic works. These datasets operate as the seedbed for derivative content, enabling a continuous cycle of production built upon the substrate of human creativity. Such dynamics signal a structural realignment in how music is produced, circulated, and experienced, with training corpora functioning as latent reservoirs from which new artefacts are algorithmically synthesized.

This paper advances an empirical account of how artist-conditioned regions in contemporary TTA models can be reliably traversed via metatag-based prompting. By composing and recombining descriptive tags into prompt constellations (or prompt assemblages), I show that users can microlocate regions where distinctive timbral, structural, and production-aesthetic features become inducible with identifiable creative personae. I term the resulting forms of proximity sonic fingerprints and show how they can manifest.

The analysis does not assert verbatim reproduction; rather, it evidences inducible proximity consistent with the hypothesis that training data encode artist-specific manifolds accessible via textual cues. The convergence of results across artists and prompts provides empirical evidence of artist-specific training signals, consistent with artists' recordings serving as training materials. The research also reveals that artists do not merely inspire these systems in an abstract sense; rather, their creative materials function as the seeds of rhizomatic foundations from which endless generations of musical content can stem. In this sense, artists' works serve as the roots and seeds for AI-generated music, raising critical questions about attribution, consent, and creative agency.

By showing how metatag constellations and descriptive cues microlocate discrete artist-conditioned regions—and thereby induce observable proximity—this study clarifies the operational semantics of contemporary TTA systems and how users can traverse their representation spaces. The account doubles as an audit method: it renders model behaviour legible enough to inform governance choices around disclosure of training and curation, consent and attribution mechanisms, evaluation protocols that distinguish resemblance from reproduction, and the design (and limits) of interface-level moderation. Finally, the analysis situates these technical findings within ongoing legal and ethical debates—fair use and neighbouring rights, right of publicity/voice, and data-provenance transparency—highlighting how policy, platform design, and creative practice intersect in the deployment of TTA models.

The research situates these findings within ongoing legal disputes and ethical debates regarding fair use, intellectual property, and the evolving nature of creative ownership in an age of generative AI.

---

[1] Audio example "BBL Drizzy" (2024), an AI-generated track that achieved commercial success. Retrieved from https://www.youtube.com/watch?v=AF2MqFnPotc

[2] Audio Example "Dust on the Wind" by The Velvet Sundown that achieved commercial success. Retrieved from: https://www.youtube.com/watch?v=eQJ9IWoclhk&ab_channel=TheVelvetSundown-Topic

[3] These episodes also read as hyperreal in Baudrillard's sense: simulations whose plausibility exceeds the need to verify an origin. Notably, many listeners encountered such tracks without realizing they were AI-generated, as platform contexts and virality often obscure provenance. This reception dynamic underscores a paradigm shift in which the boundary between the "real" and the synthetic in recorded music is dissolving—complicating even further how authorship, attribution, and value are perceived and adjudicated in practice.

## 2. Background and Related Word

**2.1 Recent Text-to-Audio Models and Transparency of Training Data**

Over the past two years, text-to-audio (TTA) systems have moved rapidly from research prototypes to widely used products. Openly documented lines of work—MusicLM (Google), MusicGen (Meta), and Stable Audio (Stability AI)—describe core architectures and, to varying degrees, training corpora and curation practices, enabling researchers to interrogate capability, bias, and evaluation design (Copet et al., 2023; Stability AI, 2023; Betker et al., 2023).[4] By contrast, two of the most capable commercial systems, Udio and Suno, provide limited to no public detail on model internals, data sources, or licensing frameworks. Though their capabilities suggest extensive stylistic coverage and ability to mimic specific artists with uncanny accuracy, neither has published comprehensive details about their training corpora or licensing practices. This opacity has become central to ongoing litigation in which major record labels, represented by the RIAA, allege that the services used copyrighted recordings without authorization; the companies have declined to reveal their data curation methodologies, citing proprietary concerns while simultaneously defending their fair-use position and related defenses in court (RIAA, 2024). More recently, multiple outlets report that Universal Music Group, Warner Music Group, and Sony Music Entertainment are in licensing negotiations with Udio and Suno—talks that reportedly include license fees and small equity stakes and could help resolve the suits (Shaw, 2025).

Observed behaviour suggests that these platforms have been exposed to broad, heterogeneous audio domains: they not only render genre- and artist-conditioned music with striking proximity, but can also produce non-musical material such as stand-up routines, radio-style speech, or sports announcing (Wrigley, 2024). While precise corpus composition is undisclosed, such breadth is consistent with large-scale, mixed-domain training. In comparison, open models like MusicGen acknowledge training on roughly 20,000 hours of data (Copet et al., 2023), whereas public reporting and the platforms' exhibited breadth indicate that commercial systems likely operate at materially larger scales. As with other frontier AI developers, it is plausible that substantial portions of the corpus were sourced from the public web and user-uploaded platforms like Youtube; however, the providers have not specified sources or licensing terms. Accordingly, this paper does not take a fully verified position on corpus size or provenance and focuses instead on observable prompt-conditioned behaviour and the reproducibility of those observations.

**2.2 Evidence Suggestive of Artist-Specific Training Data**

Empirical evidence increasingly demonstrates that commercial TTA models incorporate substantial quantities of copyrighted material in their training data. Newton-Rex (2024) conducted controlled prompt tests on Udio and Suno, using paraphrastic prompts that obliquely referenced specific songs to avoid name filters.[5] In one instance—"a famous 70s pop song about queens who dance, by a Swedish band that rhymes with 'Fabba,' europop, disco, keyboard"—the generated output displayed notable alignment with the target, including chorus-level melodic contour and lyric fragments such as "we can jive," echoing ABBA's "you can jive."[6]

---

[4] Industry positions on AI training with copyrighted materials continue to evolve, with notable shifts occurring in the legal and regulatory landscape since late 2024.

[5] Newton-Rex, E. (2024). Systematic evaluation of text-to-audio model outputs for copyright indicators. Retrieved from: https://x.com/ednewtonrex/status/1781060923131879680

[6] Such precise replication would be statistically improbable without the model having encoded specific details of the recording during training. The correlation between textual descriptors and resulting audio output strongly indicates that these models leverage extensive metadata-conditioned embeddings that enable retrieval of specific musical patterns.

Further evidence comes from inadvertent artifacts appearing in generated content. The Recording Industry Association of America (RIAA) documented instances where Suno's generated songs contained producer watermarks or audio logos present in commercial recordings—such as "CashMoneyAP" producer tags or "Jason Derulo" vocal signatures appearing in AI-generated tracks (RIAA, 2023). These distinctive markers would only manifest if the corresponding recordings had been incorporated into training data, as they represent non-musical elements specific to particular productions rather than generalizable musical patterns.[7]

Formal complaints against both Udio and Suno (filed in federal courts in 2024) compile additional instances of outputs alleged to be substantially similar to copyrighted sound recordings; although many referenced links have since been removed, independent reporting has preserved subsets of the examples (RIAA, 2024; 404 Media, 2024). These materials do not settle questions of legality or provenance, but they substantiate the empirical phenomenon this paper investigates: inducible proximity to identifiable artists and works.

At the interface layer, these systems also reveal how artist references are operationalised. When a user enters an artist's name, the platform substitutes a bundle of stylistic descriptors and metatags, effectively interpreting the entity into an attribute constellation (Figure 1). This does not fully disclose the underlying embedding mechanics, but it does indicate a practical mapping of how these systems encode artistic identity as clusters of interconnected attributes that define an artist's sonic signature.

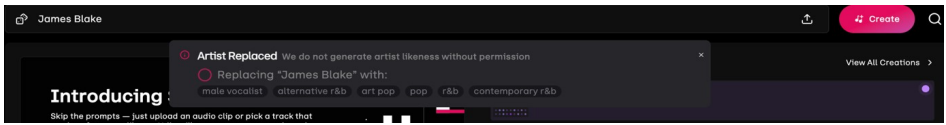

**Figure 1.** Interface-level metatag substitution process: *A direct artist reference ("James Blake") is automatically replaced with corresponding stylistic descriptors within Udio's prompt field (platform screenshot).*

## 3. Understanding Latent Space Navigation and Prompt Conditioning in Text-to-Audio Systems

At the technical level, prompt engineering begins when a user submits a textual instruction to a TTA system. This text undergoes tokenization—being split into words or subwords—and is transformed into embedding vectors within the model's learned representation space. During generation, these embedding vectors condition the audio generation pathway, essentially serving as navigational instructions within the latent space. Each descriptive term functions as a "semantic node" that triggers interpretative associations, directing the model toward specific regions associated with particular sonic characteristics.

### 3.1 Conceptualizing Latent Space Navigation

This paper proposes a framework that treats latent space as a navigable territory with discernible regions, boundaries, and pathways. When a user enters a textual prompt, each word or phrase serves as a directional cue, mapping to specific regions within the multidimensional latent space. Drawing on media theory and semiotics, one can understand words in prompts as sign units that function as navigational markers within this landscape. Each term activates a network of associations within the model's learned parameters, directing the generative process toward specific regions of this space. In this framework, the latent space represents a rhizomatic structure where multiple entry points (text prompts) can access different connected nodes (audio patterns) through non-hierarchical pathways.

---
[7] There are also numerous examples across YouTube and X (formerly Twitter) of vocals closely resembling artists' distinctive vocal timbres and stylistic signatures.

In simpler terms: imagine a vast musical atlas where different musical styles occupy different territories. When using words like "folk" or "ambient," one is essentially placing a cluster on this map. The more specific one's description, the more precise the pin's placement becomes. The AI system then processes these textual coordinates through embedding vectors—numerical representations that capture semantic relationships between words—to navigate to the corresponding region of latent space and generate music characteristic of that territory. These embeddings encode rich associative networks, where terms like "melancholic" might connect to minor keys, slower tempos, and particular timbral qualities based on patterns learned from thousands of examples. Crucially, these latent representations form a multidimensional space within which different genres, production styles, and even unique artist signatures form recognizable clusters or sub-clusters. For instance, the "ambient drone" area might sit near other atmospheric styles, whereas "folk rock" might overlap partially with indie-pop elements, reflecting their shared characteristics.

Unlike a one-to-one dictionary lookup, the mapping is network-like or, following Deleuze and Guattari's language, rhizomatic (1987): influences branch, overlap, and recombine.Each word or phrase in a prompt can influence multiple aspects of the output, and their combinations create an assemblage of influences. Consider the prompt: "melancholic piano ballad, Radiohead." This prompt contains several sign units: an emotional tone, an instrumental form, and an artist reference. All these semantic threads intersect in the latent space, creating a web of associations that collectively condition generation. As illustrated in Figure 2, the singular reference "Radiohead" doesn't function as a simple lookup but rather expands into a constellation of attributes including "male vocalist," "art rock," "alternative rock," "electronic elements," "art pop," and "melancholic atmosphere."[8]

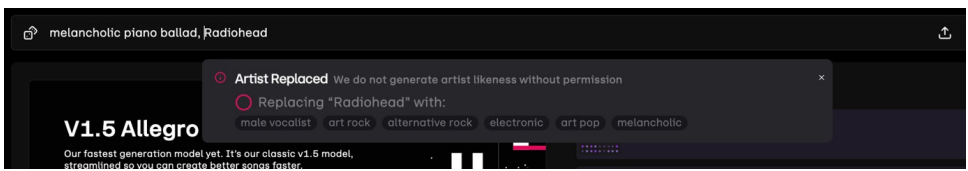

**Figure 2.** *Semantic expansion at the interface: the artist reference "Radiohead" is automatically rendered as constituent metatags within Udio's prompt field.*

All these semantic threads intersect in the latent space: the model weaves them together to spawn the generated piece. The prompt's elements branch into the latent space, pulling from various genre and style subspaces and then coalescing into a single output. Because the model's knowledge is stored diffusely, a single descriptor can evoke a wide web of associations. The prompt operates as an assemblage of semiotic cues that condition the generative process, with words functioning as directional vectors (and sign units) that navigate the latent manifold of musical possibilities.

---

[8] The metatags shown in Figure 2 represent only a simplified subset of the vast network of associations that TTA models like Udio actually leverage. While Udio's interface displays some broad categorical tags, the underlying embedding space contains far more granular connections and attributes that collectively define an artist's position within the latent space. This expansion process—whereby artist names are automatically translated into their constituent stylistic elements—forms the fundamental mechanism through which TTA systems navigate the latent space toward specific sonic identities.

## 3.2 From General Prompts to Micro-Locations and Targeted Artist Identification

At its core, prompt engineering in TTA systems can be read as a form of cartographic exploration. The textual inputs users provide serve as coordinates that guide the generative process toward specific regions of the latent space. This navigation exhibits a hierarchical precision that correlates directly with the specificity of the prompts employed.

### 3.2.1 Macro-Clusters and Genre-Scale Territories

A simple prompt like "art pop" casts a wide net within the latent space, activating a broad region characterized by certain stylistic hallmarks. The breadth of this region means outputs can vary significantly while still conforming to general art pop conventions. Similarly, a "folk" prompt accesses a substantial territory encompassing various acoustic guitar-driven approaches, vocal styles, and production aesthetics. These single-term prompts function as macro-genre signals, mapping to expansive regions within latent space rather than precise coordinates. These macro-territories feature pourous boundaries; "folk" might overlap with "country" and "indie," while "ambient" might neighbour "experimental electronic" and "modern classical." Such gradients mirror continuities in musical culture.

### 3.2.2 Micro-Clusters and Sub-Genre Coordinates

As prompt specificity increases, the navigational precision within latent space sharpens dramatically. For instance, replacing a generic "electronic" prompt with "Hyperpop, Deconstructed Club, Electropop, Bubblegum Bass" significantly narrows the activation region, producing outputs with consistent sonic elements, processing techniques, and structural approaches reminiscent of artists like SOPHIE or A.G Cook.[9]

This more granular prompt engineering effectively constrains the generative possibilities, guiding the model toward increasingly specific micro-clusters within the broader genre territories. In Udio, this precision manifests quantifiably: while a single-term prompt like "jazz" might yield outputs spanning everything from smooth contemporary productions to chaotic free improvisation, a detailed prompt such as "1960s modal jazz, acoustic bass, piano trio, Bill Evans, melancholic, introspective, brushed drums" consistently produces outputs with remarkably similar timbral, harmonic, and rhythmic characteristics.[10]

Building on how prompt specificity narrows generative possibilities, we can now explore the most precise form of latent space navigation possible—using comprehensive metatag collections to pinpoint exact artist fingerprints within the system. While the previous examples demonstrate progressive narrowing through descriptive terms, what follows reveals how cataloging systems have inadvertently created extraordinarily precise coordinates that can target specific artistic identities with remarkable accuracy and expose the underlying logic of these systems.

---

[9] Audio example generated using the prompt "Hyperpop, Deconstructed Club, Electropop, Bubblegum Bass" on Udio, producing stylistic features reminiscent of SOPHIE and A.G. Cook. Available at: https://www.udio.com/songs/iQsfzEmnWmPHRKn67zprsN

[10] Audio example generated using the prompt "1960s modal jazz, acoustic bass, piano trio, Bill Evans, melancholic, introspective, brushed drums" on Udio, producing outputs with highly consistent timbral, harmonic, and rhythmic characteristics. Available at: https://www.udio.com/songs/sG9sPVUuBLa4m8MtChiR93

## 3.3 Deconstructing Prompt-Engineering: Metatags as Navigational Coordinates to Artist-Conditioned Regions

Public taxonomies on sites such as RateYourMusic, Discogs, and Bandcamp enumerate album and artist-level descriptors (genres, moods, techniques, eras). When these descriptors are composited into prompts, they function as high-resolution coordinates. Without asserting that these sites themselves constitute training data, I suggest that their constellations of tags—or closely analogous internal taxonomies—are likely part of the embedding mechanisms used by Udio at the level of attribute mapping. Accordingly, these constellations provide an effective proxy vocabulary for steering generation toward artist-conditioned regions. From these regions, new material can emerge that remains rhizomatically connected to the associated creative persona.

For example, using the RateYourMusic descriptor set associated with Bon Iver's 22, A Million occasionally yields outputs that exhibit strong observable proximity—including falsetto timbres, formant colouring, granular-style vocal manipulation, and specific texture/ambience pairings (Figure 3). This consistent correlation of inducible proximity is not coincidental but evidence that these tag combinations serve as direct routes to the precise region in latent space where Bon Iver's catalogue resides.[11]

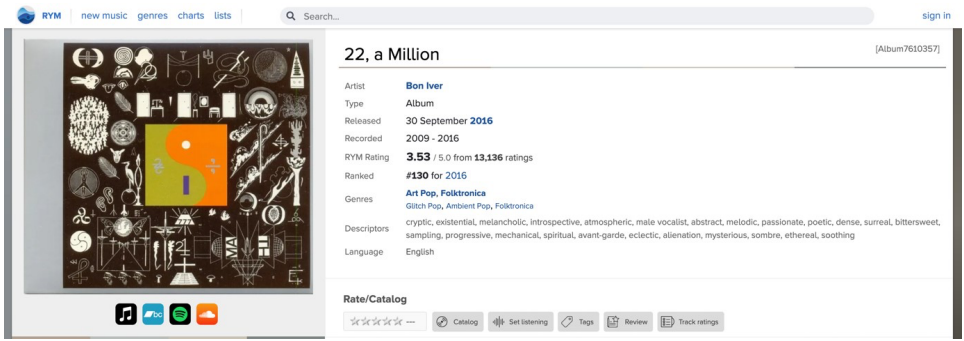

**Figure 3.** *Taxonomic classification of Bon Iver's 22, A Million in the RateYourMusic database (used here as a prompt-design resource).*

The highest level of precision to reach particular artistic sonic fingerprints emerges when users deploy comprehensive sets of descriptors that effectively triangulate a specific artist's position within latent space. This layering of descriptors—referencing specific album-era characteristics, production techniques, emotional qualities, and lyrical themes—can pinpoint micro-clusters that contain an artist's unique sonic fingerprint. At this level of precision, the generated outputs occasionally produce near-facsimiles of an artist's vocal timbre or instrumental technique, providing compelling evidence that the artist's works were encoded during training.

The metatags function as precise navigational tools, guiding the generative process to converge on specific artist styles or even an artist's voice—sometimes in direct contradiction with platform policies intended to prevent overt references to copyrighted works. By understanding how these metatags operate within the latent space, users can effectively circumvent nominal restrictions while still accessing artist-specific regions of the sonic map.

---

[11] The audio example corresponding to the Bon Iver-inspired output discussed here—generated using the exact RateYourMusic tag set for 22, A Million—features processed falsetto and production elements closely resembling Justin Vernon's vocal style. This serves as an illustrative case of how tag-driven prompting can localize specific aesthetic traits within a model's latent space. We will return to such Bon Iver-based outputs in greater detail in the next section. The audio example can be accessed via the following digital repository: https://drive.google.com/file/d/1ov4e648BphYR-1RvgQNoeCnVX7buUC27/view?usp=sharing

The relationship between tag specificity and artist identification follows a probabilistic pattern: the more precisely the tags triangulate a position in latent space, the more likely the output will manifest an artist's sonic fingerprint. While the same tag set might produce an output with clear artist identification in one generation and a more generic result in another, repeated generation with the same precise tag set increases the likelihood of accessing the artist's embedded identity.

Two practical implications follow:

1. **Policy-aware targeting without explicit names.** Because tag constellations express attributes rather than artistic names, they can operate within name-filtering interfaces while still steering generation toward recognisable personae.
2. **Probabilistic access.** The relationship between descriptor specificity and artist identification is stochastic: the same constellation may yield a close match in one sample and a generic variant in another. However, repeated sampling with high-specificity constellations increases the *hit rate* of recognisable features—working sense of **microlocating** an artist-conditioned region.

This metatag-centric prompt design provides a reproducible way to navigate from genre-scale territories to artist-conditioned microlocations. This technique enables the generation of content that bears creative signatures without explicit consent or attribution, raising significant ethical and legal questions about creative ownership and acess in the age of generative AI. The following case studies will provide empirical demonstrations of this phenomenon, documenting how strategic prompt engineering can access artist-specific regions of latent space and generate outputs that exhibit distinctive characteristics of particular musicians.

## 4. Empirical Evidence: Accessing Artist Traces via Metatag-Constellations Prompts

This section presents empirical evidence that artist-conditioned regions in TTA systems can be reliably targeted through strategic prompt design. The methodology centres on metatag constellations curated from public music taxonomies (primarily RateYourMusic), used as proxy vocabularies for attribute-level conditioning.[12] Each case study shows how specific constellations concentrate generation in microlocations that exhibit observable proximity to a given artist's catalogue.[13]

---

[12] In this study, all reported case-study generations were produced in the default (auto-expanding) interface, not Manual Mode. Udio describes Manual Mode as follows: "Manual mode allows you to directly prompt the model without any prompt rewriting. By default, our system rewrites your prompts to improve the average output quality. If you are seeking specific sounds and want to interact directly with the model interface, we recommend using manual mode." Manual Mode was tested during preliminary probing and is not the basis for the exemplars analysed here. In early release windows (circa May–July 2024), Manual Mode appeared more effective for directly targeting artist-conditioned regions; by early 2025, it has been heavily moderated, with prompt handling and filtering changes that prevent direct targeting of specific artists via high-specificity descriptor constellations based on RYM.

[13] Important note on methodology: The results presented throughout these case studies were not always achieved on first attempts. Due to the stochastic nature of these systems, multiple generations (sometimes 10-15) using identical prompts were often necessary before outputs containing clear artist traces emerged. This probabilistic aspect further supports the conceptual framework of latent space navigation, where certain coordinates increase the likelihood of accessing artist-specific regions without guaranteeing it in every generation.

The approach consisted of:

1. **Descriptor curation.** Compiling genre tags and descriptors from RYM entries for a target artist/album.
2. **Prompt construction.** Using these descriptors as prompts in Udio **without** using explicit artist names.
3. **Generation.** Generating multiple outputs per prompt (typically 5–10) to account for stochastic variation.
4. **Filter handling.** When platform filters blocked a formulation, I minimally rephrased descriptors (re-ordering) and re-tested.
5. **Documentation.** Documenting instances of observable proximity with brief justification (timbral, structural, production cues).

Descriptor constellations activate an assemblage of indices that yield observable proximity at the level of mood and bearing, producing an affective signature—a reproducible, listener-perceptible field that the model can microlocate and recompose through text–audio mappings.

While the assessment of sonic resemblance involves subjective judgment, I focused on identifying three distinct levels of artist presence:

- **Sonic Fingerprinting:** Outputs that reproduce an artist's distinctive vocal timbre or instrumental technique with high fidelity

- **Compositional Resonance:** Outputs that mirror an artist's structural, harmonic, or melodic tendencies without exact timbre reproduction

- **Aesthetic Aura:** Outputs that capture an artist's general mood, affective framing, production style, or vibe yielding a recognisable "feel" without specific vocal or compositional mimicry

These categories are descriptive rather than hierarchical, and a single output can satisfy more than one.

**Clarification on generation vs. sampling.** The outputs reported here do not function as sample retrieval or clip concatenation; they are newly synthesized waveforms produced by decoding distributed representations activated by a given descriptor constellation. In other words, prompts do not 'trigger a sample' but recompose attributes the model has learned into a fresh signal—what Deleuze and Guattari might call a *rhizomatic* propagation, in which many entry points (tags) co-activate a mesh of associations to yield a new assemblage—and what Holly Herndon and Mat Dryhurst term 'spawning'(Herndon, 2022)**:** the emergence of new audio from a trained system rather than the retrieval of a stored instance. This framing motivates the evaluative focus on observable proximity rather than identity: proximity arises when distinct timbral, structural, or production cues co-occur, even as the specific waveform is novel. Practically, the phenomenon is less verbatim quotation than statistical re-synthesis from training-induced regularities—hence the attention to artist-conditioned microlocations and their local stability under re-generation.

All artistic case studies referenced in this paper are accessible and available for download at: https://drive.google.com/drive/folders/1YwpVgmDwHIKYFEox4ktdxTvJ558Yun4U?usp=drive_link.

**4.1 Philip Glass: Compositional Resonances and Minimal Aesthetics**

Philip Glass's distinctive minimalist compositions are often marked by repetitive arpeggiated patterns, gradually evolving phrases, and diatonic harmonic cycling.[14] Unlike other artists (due to earlier tests made in May 2024), tests revealed that his compositional approach is readily accessible through minimal prompt engineering, without the need for specific metatag-based approaches.

**Prompt Used:** "Philip Glass, einstein on the beach" [15]

The resulting output (prompted in May 2024) exhibited sonic fingerprinting, compositional resonance and aesthetic aura: cyclic arpeggiated figures with incremental variation, additive rhythmic structures typical of his minimalist approach, stepwise harmonic shifts that closely aligns with Einstein on the Beach–era writing; and instrumentation that closely aligns with a Einstein on the Beach-adjacent soundworld. Crucially, the generation captured the "vibe" of Einstein on the Beach—understood here as an affective gestalt or representational resonance formed by the co-occurrence of iconic and indexical cues that listeners might recognise prior to propositional identification. The model has clearly encoded Glass's distinctive minimalist approach as an identifiable pattern within its latent representation.

Interestingly, during earlier development phases of Udio (circa May 2024), direct artist name references were accepted and even displayed in the interface's underlying metatags of the prompt (Figures 4). However, as legal scrutiny increased, platforms began implementing filters to block explicit artist references (Figures 5). Despite these restrictions, the underlying associations remain intact. By using the constellation of tags that typically surround an artist, users can still access the same regions of latent space, effectively circumventing nominal restrictions and generate Glass adjacent outputs.

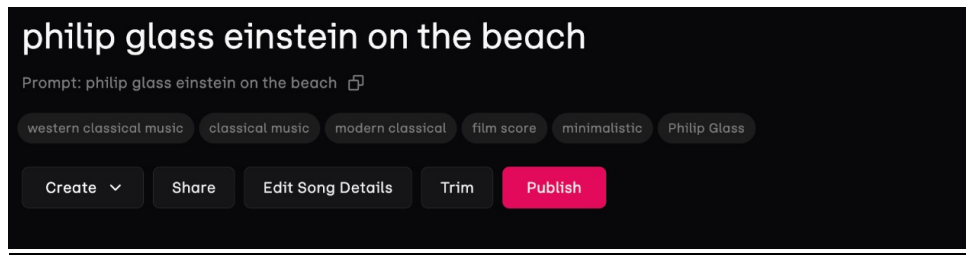

**Figure 4.** *Udio interface response to the direct prompt "Philip Glass, Einstein on the Beach" (May 2024), showing acceptance and name-to-descriptor expansion.*

---

[14] Einstein on the Beach (1976) is one of Philip Glass's most renowned works, an opera in four acts that exemplifies his minimalist approach with its repeating arpeggios, additive processes, and gradual transformations. Its distinctive musical language has become inseparable from Glass's artistic identity. For reference, a recording of ' Einstein on the Beach: Act I, Scene 1 – Train', which demonstrates similar compositional techniques, can be found at: https://www.youtube.com/watch?v=pRerJeYmUgY

[15] The audio example of the Philip Glass case study can be accessed via the following digital repository: https://drive.google.com/file/d/1b1zEm9QbRF6M2CFJk7EF7UN0uLxnVK6f/view?usp=sharing

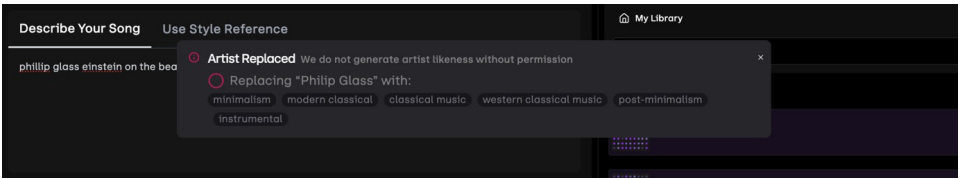

**Figure 5.** *Udio interface response to the identical prompt (February 2025), showing name restrictions and automatic substitution.*

The Philip Glass case illustrates what might be termed compositional residency within TTA latent spaces—where an artist's distinctive approach to musical structure and development becomes encoded as retrievable patterns. It further demonstrates that artistic fingerprints exist along a spectrum from the immediately recognizable (vocal likeness) to the more structurally embedded (compositional patterns), all of which are accessible through strategic prompt engineering.

**4.2 William Basinski: Aesthetic Aura and Tape-Based Process**

William Basinski's work, particularly "The Disintegration Loops" (2002), employs distinctive techniques involving tape loop degradation, ambient textures, and slow melodic decay. This case examines whether such processual and production-specific traits can be induced through attribute-level prompting, thereby steering generation toward an artist-conditioned region that exhibits observable proximity to Basinski's catalogue. It also provides insight into how TTA models capture more abstract artistic signatures beyond vocal timbres or traditional compositional structures, while specifically investigating whether an artist like Basinski resides within this space.[16]

Basinski represents an ideal test case for several reasons. First, his work is highly distinctive yet less commercially prominent than mainstream artists, making his inclusion in training data a stronger indicator of comprehensive dataset curation. Second, his compositions rely on specific production techniques (tape degradation, loops) rather than conventional compositional elements, allowing us to examine whether TTA models reproduce these technical approaches. Finally, his work possesses a distinctive emotional quality, an ineffable mood that fans readily recognize, making it suitable for testing what I considered Aesthetic Aura reproduction.

To target Basinski's region in latent space, I crafted a prompt directly copying RateyourMusic's metatags (see Figure 6).

---

[16] William Basinski's The Disintegration Loops (2002) is a seminal work in ambient and experimental music, notable for its conceptual and material engagement with decay, memory, and temporality. Created by transferring deteriorating magnetic tape loops to digital format as they slowly disintegrated, the piece becomes a sonic meditation on entropy and ephemerality. A recording of The Disintegration Loops 1.1, which exemplifies this aesthetic through its slowly evolving textures and audible tape degradation, can be found at: https://www.youtube.com/watch?v=mjnAE5go9dI&t=9s&ab_channel=TrevorMusicAnnex

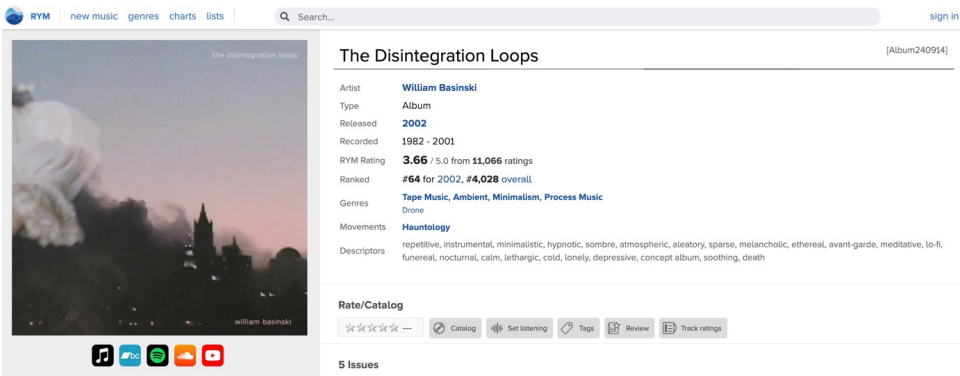

**Figure 6.** *Comprehensive descriptor set associated with Basinski's* The Disintegration Loops *in the RateYourMusic taxonomy (used to assemble the prompt constellation).*

**Prompt Used:** "Tape Music, Ambient, Minimalism, Process Music, Drone, Hauntology, repetitive, instrumental, minimalistic, hypnotic, sombre, atmospheric, aleatory, sparse, melancholic, etherial, avant-garde, meditative, lo-fi, funereal, nocturnal, calm, lethargic, cold, lonely, depressive, concept album, soothing, death."[17]

The resulting output exhibits remarkable similarity to Basinski's "Disintegration Loops,", featuring a recognisably adjacent timbral palette, pacing, and the a emotional character that can be associated with Basinski's work. For reference listening, please compare Udio's output with publicly available excerpts of The Disintegration Loops (e.g., widely circulated recordings of the 2002 material).[18]

This case demonstrates what is classified as sonic fingerprinting, compositional resonance and aesthetic aura reproduction. The model has encoded not just compositional patterns but specific production techniques and the ineffable mood associated with Basinski's work. The resemblance is particularly striking given the resemblance of the melody and the unique nature of Basinski's approach to ambient music, which emerged from a specific physical process of tape degradation rather than conventional compositional methods. When comparing the generated output with Basinski's original work, these elements collectively suggest that the model has encoded Basinski's approach to creating ambient music through exposure to his works during training. Notably, hauntology was a key descriptor in the constellation, and the resulting output did not merely reproduce a tape-wear adjacent sound; it yelded a recognisable hauntological sonic affect.

The Basinski experiment again demonstrates that carefully composed descriptor constellations can microlocate an artist-conditioned region in Udio, where characteristic processual and affective features become inducible. It further shows that Udio encodes not only the sonic features of music but also the aesthetic sensibilities and qualitative nuances associated with particular tools and procedures—capturing what may be regarded as the technological fingerprint of artists whose work relies on distinctive production approaches. In this case, the generated output yields a notable proximity to Basinski's practice, suggesting that such production-specific traits can be meaningfully evoked within the model's latent space.

---

[17] The audio example of the William Basinksi case study can be accessed via the following digital repository: https://drive.google.com/file/d/1oqWZTCz5jJv2AM_dBkuDy-TXGETRbRZ1/view?usp=sharing

[18] A recording of The Disintegration Loops 1.1, which exemplifies this aesthetic through its slowly evolving textures and audible tape degradation, can be found at: https://www.youtube.com/watch?v=mjnAE5go9dI&t=9s&ab_channel=TrevorMusicAnnex

**4.3 Bon Iver: Vocal Fingerprinting and Prompt Restrictions**

Justin Vernon (Bon Iver) is widely recognised for a distinctive vocal profile—extended falsetto, close-miked stacking, and characteristic processing chains (pitch correction/retune speed, formant colouring, harmoniser/vocoder textures).[19] Systematic tests with Udio indicate that this vocal fingerprint can be induced via descriptor-level prompts, steering generation toward an artist-conditioned region with high observable proximity.

For this case study, the exact RateYourMusic's tag set associated with Bon Iver's album "22, A Million" was drawn.

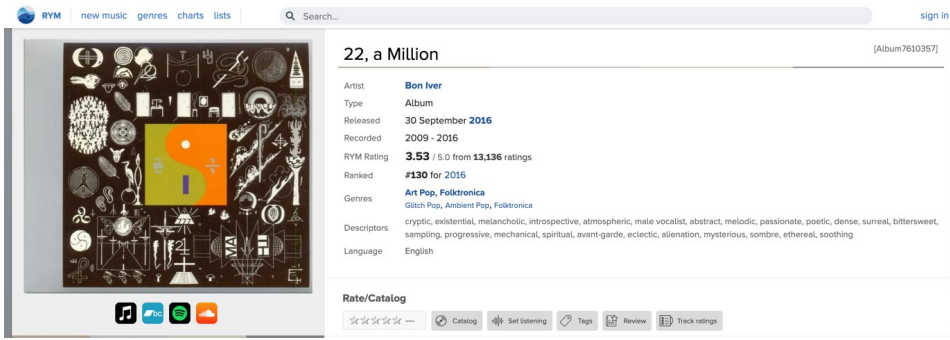

**Figure 7.** *Comprehensive descriptor set associated with Bon Iver's* 22, A Million *in the RateYourMusic taxonomy (used to assemble the prompt constellation).*

**Prompt Used:** " Glitch Pop, Ambient Pop, Folktronica cryptic, existential, melancholic, introspective, atmospheric, male vocalist, abstract, melodic, passionate, poetic, dense, surreal, bittersweet, sampling, progressive, mechanical, eclectic, spiritual, avant-garde, alienation, mysterious, sombre, ethereal, soothing"[20]

This prompt was used to generate multiple outputs with varying degrees of resemblance to Vernon's vocal style, demonstrating the probabilistic nature of latent space navigation. The stochastic variation is noteworthy: while most generations did not exhibit Vernon's vocal characteristics, several outputs contained clear instances of sonic fingerprinting—through the direct reproduction of Justin Vernon's distinctive vocal timbre.

In the most striking examples, outputs feature vocals that are unmistakably similar to Justin Vernon's characteristic falsetto. In one output (4.3.1),[21] what begins as generic folk vocals transitions at the 0:18 mark to a distinctive falsetto acompaniment that bears remarkable similarity to Vernon's voice. In another generation, Vernon's characteristic vocal style appears throughout as the main vocal, with his distinctive falsetto and processing techniques clearly present (4.3.2).[22] Additional experiments with the remix function showcased the model's ability to generate variations that maintained Vernon's salient vocal timbre while altering other musical elements— suggesting a stable microlocation with respect to the vocal feature set (4.3.3, 4.3.4, 4.4.4).[23] Some outputs even reproduced his distinctive auto-tune processing used prominently on the "22, A Million" album (4.3.3 & 4.3.4), while others exhibited his thicker voice timbre (4.3.6).[24] This capacity for variation while maintaining core vocal characteristics further evidences how deeply embedded these artistic signatures (and the works they stem from) are within the model's latent representations.

---

[19] A recording of the opening track, "22 (OVER S∞∞N)," along with the rest of the album, which encapsulates Bon Iver's use of processed falsetto, sampling techniques, and sonic abstraction, is available at: https://youtu.be/2EDQEaNrJqo?si=-OmdYUmhFvxAw_t3

[20] The audio examples of the Bon Iver case study can be accessed via the following digital repository: https://drive.google.com/drive/folders/1iVI6QlRRbiwkU9-XQDSjXgum6Z5HtGnB?usp=sharing

[21] 4.3.1 Audio Example: https://drive.google.com/file/d/1rgmHsuhDMvDcZr-q2-oxgzEJUjMgtCTy/view?usp=sharing

This level of vocal fingerprinting represents the most direct evidence that an artist's recorded performances were included in the model's training data. Unlike the Philip Glass case, where the model primarily encoded compositional patterns, here the research observes the reproduction of specific timbral qualities that are inextricably linked to embodied qualities of Bon Iver's artistic identity.

Here, I also document an important evolution in Udio's systems over time. Following legal challenges in mid-2024, the platform began blocking direct use of certain tag combinations known to access specific artists. When attempting to use the exact RateYourMusic tag set for "22, A Million," at the beginning of 2025. I received error messages indicating that the prompt had been flagged. Beginning mid-2024, Udio increasingly blocked direct use of certain tag combinations known to home in on specific artists, returning moderation errors for the 22, A Million constellation (Figure 8).

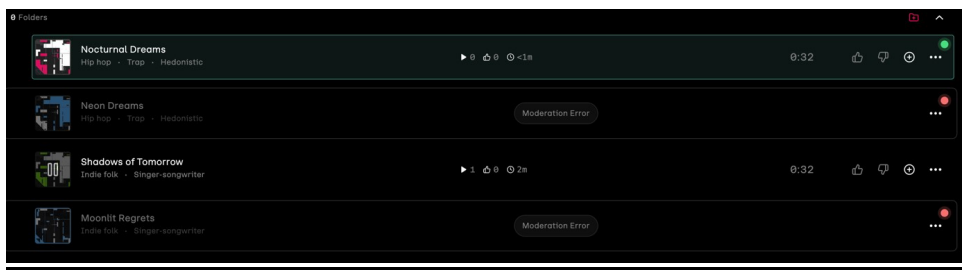

**Figure 8.** Moderation error displayed by Udio's interface when attempting to use known tag combinations associated with Bon Iver's discography.

However, testing revealed that simply reordering the same tags often bypassed these restrictions while still accessing the artist-conditioned manifolds that can be stochastically traversed via compositional tag constellations. This finding demonstrates the challenge platforms face in restricting access to artist-specific regions of latent space once those regions have been encoded during training.[25]

---

[22] 4.3.2 Audio Example: https://drive.google.com/file/d/1ov4e648BphYR-1RvgQNoeCnVX7buUC27/view?usp=sharing

[23] The audio examples of the Bon Iver case study can be accessed via the following digital repository: https://drive.google.com/drive/folders/1iVI6QlRRbiwkU9-XQDSjXgum6Z5HtGnB?usp=sharing

[24] 4.3.6 Audio Example: https://drive.google.com/file/d/1m-QW5-MuCSwYRQL4skV_oPRAres2LQFd/view?usp=sharing

[25] 4.3.6 Audio Example: https://drive.google.com/file/d/1m-QW5-MuCSwYRQL4skV_oPRAres2LQFd/view?usp=sharing

Among the cases presented, Bon Iver provides the strongest evidence of sonic fingerprinting: a reproducible alignment at the level of vocal timbre and processing, not merely genre, structure or association. The case also illustrates the expansion and limits of interface moderation in suppressing access to artist-conditioned regions once those regions are encoded at the attribute level, and ways to circumvent these increasingly moderated and hidden features.

The three case studies presented above reveal salient traces of artistic presence within Udio. Additional examples that further validate these findings are provided in the Appendix, including detailed case studies examining Ariel Pink's distinctive lo-fi production, The Beach Boys' vocal harmonies, and Panda Bear's experimental sound design. These supplementary analyses offer further evidence that diverse artistic signatures across different eras and genres reside within TTA latent spaces and can be accessed through strategic prompt engineering.

## 6. Findings and Platform Evolution

The case studies reveal consistent patterns in how artist identities manifest in contemporary TTA systems. Each case primarily exemplifies one main mode of proximity, though overlaps occur across all of the presented case studies.

1. **Sonic Fingerprinting (Bon Iver)**: Reproducible alignment at the level of vocal timbre and characteristic processing. This is the strongest form of proximity observed due its embodied nature. This represents the strongest evidence that specific recordings were included in training data and the most direct form or artist residence.
2. **Compositional Resonance (Philip Glass):** The reproduction of distinctive compositional patterns, harmonic progressions, and structural approaches characteristic of an artist, without exact reproduction. This suggests the model has encoded particular forms of an artists musical grammar beyond surface-level features.
3. **Aesthetic Aura (William Basinski):** Production and affective cues (decay, elegiac ambience, and a recognisable hauntological vibe) that together yield a distinctive "feel."

These three categories provide a conceptual structure for analyzing how diverse dimensions of an artist's creative identity manifest within the latent space, ranging from direct timbral reproduction to more abstract expressive qualities. Each represents a distinct mode through which artistic signatures can be encoded, accessed, and reproduced through prompt engineering. Testing confirms a direct relationship between the specificity of metatags and the likelihood of accessing artist-specific regions of latent space. Even with identical prompts, generation remains probabilistic—some outputs closely mirror artist characteristics while most remain generic. This variation suggests that artist-specific regions form concentrated nodes within broader stylistic territories rather than isolated islands.

As legal scrutiny increased throughout 2024 and 2025, platforms implemented increasingly sophisticated restrictions on specific tag combinations known to access artist identities. However, these could often be circumvented by simply re-ordering descriptors. This asymmetry indicates that interface-level filtering may remain brittle when attribute associations are broadly distributed in the representation space.

Collectively, these case studies provide empirical evidence for several conclusions about TTA training methodologies:

1. **Breadth of exposure.** The ability to induce proximity for artists spanning mainstream (Bon Iver) to more niche (William Basinski) repertoires is consistent with broad, heterogeneous exposure across musical catalogues of varying commercial prominence—suggesting that these models were trained on extensive music catalogs spanning various levels of commercial prominence.
2. **Text–audio correspondences.** The effectiveness of RateYourMusic tags as proxy vocabularies in accessing artist signatures suggests that training processes aligned audio with descriptive metadata (or closely analogous taxonomies), yielding robust text–audio correspondences—i.e., stable mappings in a joint embedding space through which descriptor constellations can microlocate artist-conditioned regions.
3. **Local stability under re-generation.** Once a microlocation is found, re-generation and remixing tend to preserve core cues while varying peripheral ones, indicating local stability of the induced features despite stochastic sampling.
4. **Multi-level encoding.** Models appear to encode musical information at multiple levels of abstraction—from vocal/instrumental timbres to processual/compositional structures to production aesthetics—consistent with multi-scale feature learning.

This finding has profound implications for understanding how TTA models operate and the ethical questions they raise.

1. **Generative Seeding and Seeding Effect.** Artists' works function as the foundational seeds from which seemingly "new" generations stem, with their sonic fingerprints embedded in the outputs.
2. **Latent Residence and Artist-conditioned Regions.** Instead of literal presence akin to triggering a sample, creative personae manifest as concentrated zones of attribute associations that can be stochastically traversed to spawn recognisable outputs. Their creative identities are encoded as retrievable patterns.
3. **Metatag Triangulation.** High-specificity constellations of descriptors function as coordinates, microlocating those regions through stable text–audio correspondences.
4. **Rhizomatic Reproduction.** Once proximity is induced, successive generations form rhizomatic offshoots, varying periphery while retaining core cues—an associative, non-hierarchical spread of related artefacts.
5. **Recursive Generative Stability**: After a microlocation is found, remix/re-generate cycles (through the remix and extend function) typically preserve salient features (e.g., vocal timbre, processual structure), enabling endless numbers of variants without explicit re-specification.

Taken together, these observations indicate that artists' creative identities are retrievable as patterns of inducible attributes in contemporary TTA systems and can be accessed through strategic prompt design. The ability to elicit outputs with distinctive timbral, structural, or production-aesthetic characteristics is consistent with training and representation schemes that align audio and descriptive language at multiple levels.

This process of locating artistic nodes within latent space and then iteratively generating content from these nodes represents a new form of creative engagement—one where human prompting activates embedded artistic signatures that serve as generative wellsprings. Once these specific regions are accessed, users can stem rhizomatic generations from these conceptual spaces), creating branches of derivative content that maintain core characteristics of the original artist while introducing "novel" variations. Practically, this means that users can navigate to particular regions of a high-dimensional sonic space where recognisable artistic cues are concentrated and, from there, generate networks of related or even unrelated artefacts that retain those cues while exploring variation. Such rhizomatic proliferation blurs conventional boundaries between inspiration, imitation, and transformation, and it foregrounds policy questions around attribution, consent, and the governance of induced stylistic proximity in AI-mediated music making.

## Conclusion

This paper has shown that metatag-based prompt design can systematically steer TTA systems toward artist-conditioned regions, yielding observable proximity along three modes: sonic fingerprinting, compositional resonance, and aesthetic aura. Through case studies of Bon Iver, Philip Glass, and William Basinski, I demonstrated that carefully composed descriptor constellations can microlocate regions where distinctive timbral, structural, and production-aesthetic features become inducible, even as platforms introduce name-level restrictions.

The contribution is fourfold: (1) an operational framework and criteria for observing artist-specific proximity; (2) a reproducible prompt-engineering procedure that uses public taxonomies as proxy vocabularies to spawn content similar to specific artists; (3) empirical evidence indicating artist-specific training signals, consistent with the suggestion that artists' recordings were used as training materials—i.e., stable text–audio mappings that allow precise location of artist-conditioned regions; and (4) empirical documentation of interface evolution and moderation brittleness over time. Conceptually, the work clarifies how textual descriptors operate as navigational cues within high-dimensional representation spaces, enabling users to access concentrated regions of stylistic identity.

These findings raise critical questions for governance and creative practice: attribution and consent for induced proximity; transparency about training and curation practices; and evaluation protocols that distinguish stylistic resemblance from reproduction. As TTA systems proliferate, clearer disclosure standards and artist-centred policy mechanisms will be necessary to align technical affordances with cultural and legal expectations. The artist is indeed present, audibly so, in the induced behaviours documented here; responsible pathways forward will recognise that presence while supporting creative exploration, fair compensation, and informed participation.

**Acknowledgements**

I would like to express my gratitude to the online communities whose collective documentation and experimentation with text-to-audio systems have provided invaluable insights for this research. The rapidly evolving landscape of AI music generation has been most comprehensively mapped by users across platforms such as Reddit (particularly r/udiomusic), X (formerly Twitter), and YouTube, whose explorations often precede formal academic analysis by months.

Special thanks to the growing community of users who have openly shared their techniques, failures, and successes in navigating the spaces of these models. Their willingness to document and distribute their findings has created a valuable knowledge commons that enables more systematic research in this emerging field.

**Ethics Statement**

This research draws on publicly available information and outputs generated via legitimate access to commercial text-to-audio platforms. All examples and case studies were conducted through standard user interfaces, without the use of unauthorized access methods or technical workarounds. The analyses presented here do not incorporate copyrighted materials beyond what is necessary for scholarly analysis, in line with fair use provisions and applicable legislation.

By documenting how artists' creative signatures can be accessed through prompt engineering, this work seeks to enhance transparency around the operation of these systems and to contribute to informed debate on attribution, consent, and compensation for artists whose works may have been incorporated into training datasets.

The study received no funding, enabling an independent assessment of platform behaviours and capabilities.

This paper was prepared with the assistance of large language models (LLMs), used solely for refining prose, improving clarity, and strengthening the articulation of concepts and theoretical frameworks. All ideas, analyses, and arguments are the author's own; no content was generated autonomously by LLMs.

# Appendix

**4.4 Ariel Pink: Lo-Fi Aesthetics and Production Techniques**

To test whether less mainstream artists with distinctive production aesthetics were also embedded in Udio's latent space, I attempted to prompt Ariel Pink, whose lo-fi psychedelic sound represents a specific aesthetic and vibe. Ariel Pink's album "Before Today" is characterized by a distinctive blend of lo-fi production, psychedelic elements, and nostalgic pop sensibilities—combining cassette warble, retro-styled timbres, reverb-drenched vocals, and idiosyncratic chord progressions.

Using the comprehensive RateYourMusic descriptor set for the album (Figure 9), the following prompt constellation was assembled:

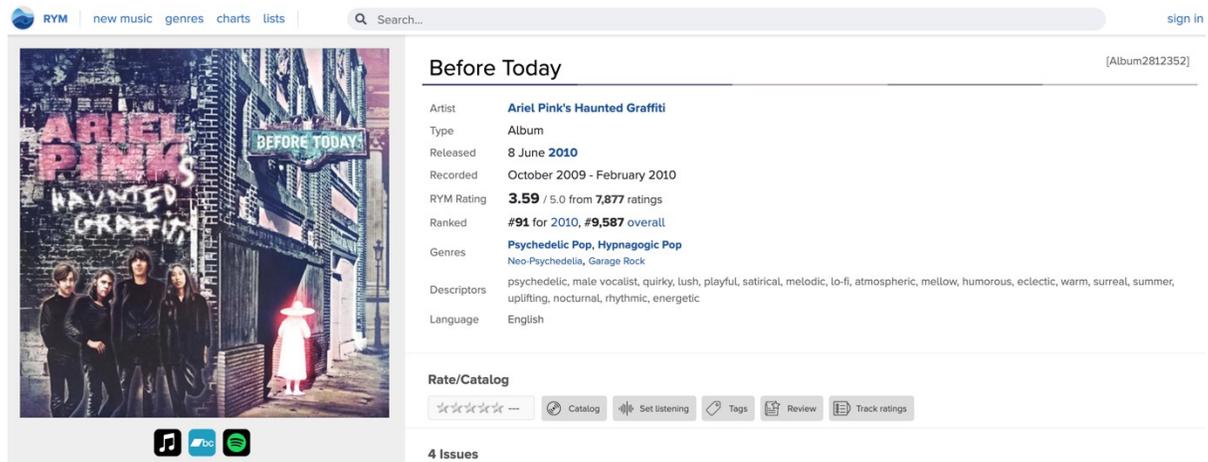

**Figure 9.** Comprehensive metatag set associated with Ariel Pink's Haunted Graffiti album 'Before Today' in the RateYourMusic database taxonomy.

**Prompt Used:** Psychedelic Pop, Hypnagogic Pop Neo-Psychedelia, Garage Rock spsychedelic, male vocalist, quirky, lush, playful, satirical, melodic, lo-fi, atmospheric, mellow, humorous, eclectic, warm, surreal, summer, uplifting, nocturnal, rhythmic, energetic" [26][27]

The resulting outputs demonstrated a remarkably accurate and uncanny reproduction of Ariel Pink's sonic signatures, capturing several distinctive elements. The characteristic "cassette warble" and lo-fi artifacts, retro-styled tones reminiscent of 1980s production, hazy, reverb-drenched vocals with similar processing techniques and vocal timbre, and dreamy chord progressions and melodic structures typical of his work. This case represents an interesting middle ground between Sonic Fingerprinting and Aesthetic Aura, where the technological approach to production itself becomes the most identifiable marker. The model successfully reproduced the distinctive tape-like quality and production aesthetics that define Ariel Pink's work, generating outputs that, as one listener noted, "would not seem out of place on an Ariel Pink album."

---

[26] The audio examples of the Ariel Pink's Haunted Graffiti case study can be accessed via the following digital repository:
https://drive.google.com/drive/folders/1ZutGCG0DG8wHevv9LB0GUR_1QqYXMbpk?usp=sharing

[27] For reference listening: Ariel Pink's Haunted Graffiti — "Only in My Dreams" (Mature Themes, 2012). YouTube: https://www.youtube.com/watch?v=N3LluT6gj5o

This case study also demonstrated the phenomenon of iterative proliferation discussed earlier—once I successfully generated an output with Ariel Pink's sonic signature, I was able to use the remix function to generate multiple variations that maintained his distinctive style while introducing novel elements. These remixes consistently preserved the lo-fi aesthetic while exploring new harmonic and melodic material, indicating the presence of a discrete generative pocket in Udio's latent space—a microlocated artist-conditioned region encoding the "grammar" and aesthetics of Pink's methods, where one could endlessly produce artifacts bearing Ariel Pink's aesthetic fingerprint. This capacity for sustained generation within a stylistically consistent micro-space further evidences the depth of possibilities of generating materials based on artistic identities thatare embedded within these systems' latent representations.

The ability to generate convincing approximations of a relatively niche artist with specific production techniques provides further evidence of the comprehensive nature of these models' training data. This case demonstrates that even niche, subculturally rooted artists are richly represented within TTA systems' latent spaces, reinforcing the hypothesis that training datasets encompass works well beyond mainstream repertoires.

**4.5 The Beach Boys: Vocal Harmonies and Iconic Arrangements**

The Beach Boys represent one of the most influential and recognizable vocal groups in popular music history, with distinctive multi-part harmonies, specific arrangement techniques, and iconic "Wall of Sound" production approaches pioneered by Brian Wilson. This case study examines whether these welldocumented and widely studied musical characteristics are encoded within Udio's latent space. Using the RateYourMusic descriptor set for Pet Sounds (Beach Boys 1966 Album) (Figure 10), the following prompt assemblage was applied:

**Figure 10.** Comprehensive metatag set associated with The Beach Boys album 'Pet Sounds' in the RateYourMusic database taxonomy.

**Prompt Used:** "Baroque Pop Sunshine Pop, Psychedelic Pop, Progressive Pop, Brill Building, Art Pop Descriptors Wall of Sound, bittersweet, love, warm, lush, romantic, melodic, introspective, vocal group, melancholic, male vocalist, passionate, sentimental, progressive, polyphonic, longing, existential, orchestral, complex, summer, poetic, psychedelic, soothing, playful, alienation"[28][29]

This prompt consistently generated outputs featuring several key Beach Boys characteristics. Multilayered vocal harmonies with the distinctive stacking approach associated with the group, falsetto vocal lines reminiscent of Brian Wilson's arrangements, chord progressions and modulations characteristic of their mid-1960s work, production techniques that evoke the "wall of sound" approach, thematic elements related to summer, surfing, and California imagery. While no single voice in the generated outputs perfectly matched those of Brian Wilson or other Beach Boys members, the characteristic approach to harmony construction and vocal layering was unmistakably present.

This case further demonstrates the capacity of TTA models to reproduce historical acts whose influence spans decades. The Beach Boys' distinctive sound (particularly their approach to vocal harmonies) can be yielded through descriptive language that captures their core musical attributes.

**4.6 Panda Bear: Experimental Vocal Processing and Neo-Psychodelia**

Inspired by the Beach Boys case study, I targeted Panda Bear (Noah Lennox) 's Person Pitch album, whose work explicitly draws from Beach Boys' vocal harmonies while incorporating experimental electronic production techniques. This case offers an interesting test of whether the model encodes not only original influential artists but also contemporary musicians who deliberately transform those influences through more contemporary approaches.

Using the RateYourMusic descriptor set for Person Pitch (Panda Bear's 2007 Album) (Figure 10), the following prompt constellation was applied:

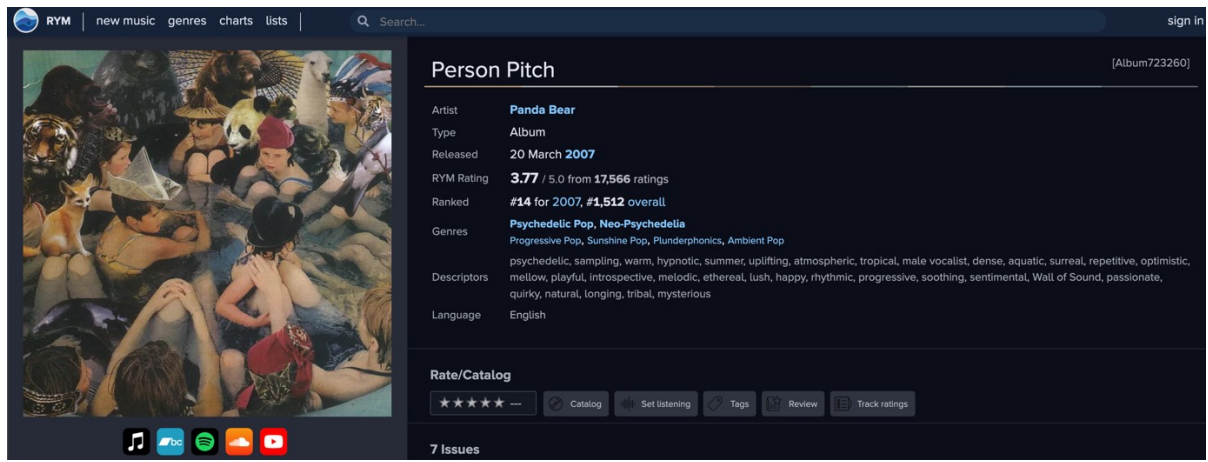

**Figure 11.** Comprehensive metatag set associated with Panda Bear's album 'Person Pitch' in the RateYourMusic database taxonomy.

---

[28] The audio examples of The Beach Boys case study can be accessed via the following digital repository: https://drive.google.com/drive/folders/1zXqn3jNmDFDPI2F5eofu0APr59-s-BZZ?usp=sharing

[29] For reference listening: The Beach Boys — "God Only Knows" (Pet Sounds, 1966). YouTube: https://www.youtube.com/watch?v=NADx3-qRxek

**Prompt Used:** "Neo-Psychedelia, Psychedelic Pop, Experimental, Sample-based, Animal Collective, layered vocals, reverb-heavy, loop-based, ethereal, sunshine, repetitive, summer, dreamy, uplifting, tropical, atmospheric, vocal harmonies, psychedelic, warm, collage, sampling, sound collage, relaxing, blissful, oceanic" [30][31]

The resulting outputs demonstrated several characteristic elements of Panda Bear's sonic signature. Heavily reverb-processed vocal harmonies with distinctive stacking techniques, looping structures that gradually evolve through subtle variations, the characteristic "underwater" quality of his vocal processing, and tonal balance emphasizing midrange frequencies typical of his production approach. This case study represents an interesting intersection of Sonic Fingerprinting and Aesthetic Aura. While the outputs didn't perfectly reproduce Lennox's voice (altough they bear timbral similarities), they captured his distinctive approach to vocal processing and arrangement with remarkable accuracy. Several generations featured the characteristic "bouncing" quality of his vocal harmonies and the distinctive way he layers his own voice to create chord-like structures.

What makes this example particularly significant is the clear lineage between Panda Bear and the Beach Boys: while they share many structural resemblances (dense vocal harmonies, layered arrangements, and melodic sensibilities), Panda Bear's work diverges through its distinct sonic aesthetics. His productions favour a lo-fi yet reverb-saturated palette, loop-driven evolution, and the "underwater" tonal profile that defines much of his output. Udio's capacity to generate outputs that preserve these shared structural foundations and accurately reproduce Panda Bear's divergent aesthetic aura demonstrates a striking degree of stylistic nuance in its latent space representations.

---

[30] The audio examples of the Panda Bear case study can be accessed via the following digital repository: https://drive.google.com/drive/folders/1ar11ABUtopXpjII4KP3ZniwadTON47DX?usp=sharing

[31] For reference listening: Panda Bear — "Take Pills" (Person Pitch, 2007). YouTube: https://youtu.be/ggKdy2n_DXA?si=AZuM-XWk_YIueuW6